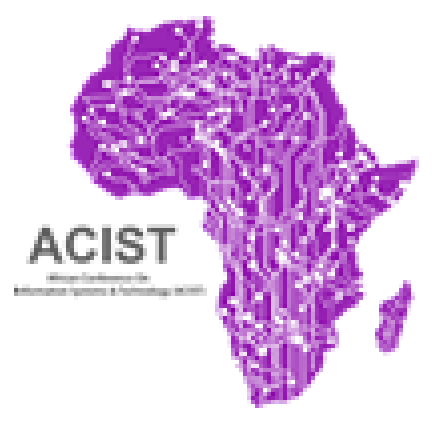

# Predicting Agile Success: The Critical Few Factors

**Ridewaan Hanslo**
University of Cape Town
ridewaan.hanslo@uct.ac.za

**Maureen Tanner**
University of Cape Town
mc.tanner@uct.ac.za

## ABSTRACT

While Agile projects are more successful than traditional software development project management approaches, their overall success rate remains relatively low, with a high percentage of projects still deemed challenged or failed. This low project success rate is attributed to projects not being rigorously evaluated against critical success factors (CSF) of Agile projects and contemporary project success criteria. To identify the CSF that contribute to Agile software development project success as perceived by Agile practitioners, this study used a positivist approach to investigate the CSF of Agile software development projects. PLS-SEM with SmartPLS was used to analyse the data and test the hypotheses to identify significant relationships between the constructs and project success criteria. This research found that a few critical factors significantly contribute to Agile project success. The study developed a novel model that can be used to evaluate and measure project success.

## INTRODUCTION AND BACKGROUND

Software development can transform ideas and concepts into software solutions. However, initially, software development had limited project management processes, leading to high risks of failed outcomes (Brooks, 1995). Traditional Software Development Methods (TSDMs), such as the Spiral, Unified Process, and Waterfall models, provided structure but faced challenges like heavy documentation and difficulty in making requirements changes (Boehm, 2002). Consequently, the Agile Manifesto was developed to establish a more adaptable and efficient approach (Beck et al., 2001). Across multiple studies, Agile software development projects have been reported as being more successful than the rigid TSDM methodologies (Hanslo & Tanner, 2020).

Despite the higher success rate of Agile projects compared to TSDMs, most projects are considered challenged or failed (Hanslo & Tanner, 2020). Studies suggest that approximately 50% of Agile projects are challenged, and 10% are failures, creating a huge global financial burden (Hanslo & Tanner, 2020). Research has applied statistical modelling to identify Critical Success Factors (CSF) as the few key areas that must go right for desirable outcomes, posited to improve this status quo (Rockart, 1979).

However, there are notable inconsistencies when reviewing Agile project success studies. First is the disparity between the studies on critical factors. While there is overlap, there are equal, if not more, variations in factor significance. For example, Stankovic et al. (2013) and Chow and Cao (2008) share one factor of significance in common, while five are different. A second inconsistency emerges from the varying number of CSF identified, which contradicts the fundamental principle of CSF theory emphasising the "few" key areas crucial for organisational success. Finally, there is a lack of research from under-represented regions, such as Africa, as most existing research is concentrated in North America and Europe (Chow & Cao, 2008; Nguyen, 2016). Including under-represented regions can reveal novel insights and uncover CSF overlooked in existing literature.

This study seeks to address inconsistencies and deficiencies in prior research. While a previous study (Hanslo & Tanner, 2026) utilised a systematic literature review (SLR) to synthesise factors from the literature (thereby identifying key factors), this conference paper specifically addresses the following research question: Which critical success factors, as perceived by Agile practitioners, have a significant relationship with Agile software development project success?

To investigate this, a global survey is conducted to include diverse representation, with the goal of including Agile practitioners from under-represented regions. Partial Least Squares Structural Equation Modelling (PLS-SEM) is employed to test the hypothesised relationships between the identified CSF and Agile project success. From a practical perspective, the few significant critical factors articulated more clearly in this study should assist stakeholders in understanding what is required from them to improve their Agile software development project outcomes.

## THEORETICAL FRAMEWORK

While there is not one single dominant theory universally applied to explain software development project success and failure, several theories are commonly used, such as the CSF, Socio-Technical Systems (STS), Complex Adaptive Systems (CAS), and Contingency theories. Within the context of this study, CSF is the primary relevant theory included as part of the Theoretical Framework (TF) to help address the research question and objective.

CSF theory provides a theoretical lens for understanding and measuring the factors underpinning organisational success. The CSF theory created by Rockart (1979, p. 9-10) is defined as: "The limited number of areas in which results, if they are satisfactory, will ensure successful competitive performance for the organisation. They are the few key areas where 'things must go right' for the business to flourish.". CSF are areas of activity that should receive constant and careful attention from management, requiring the status of their performance to be continuously measured, with the provision of information availability. After critically evaluating various theories, the CSF theory emerged as the ideal primary framework for answering the research question.

### Application of the CSF Theory as the Primary Theoretical Framework

As detailed in the online supplementary material (Appendix H), the Agile Project Success Theoretical Framework (APSTF) was developed based on a SLR (Hanslo & Tanner, 2026) and CSF theory. This framework serves as the theoretical basis for our investigation.

**Agile Project Success Criteria (Dependent Variable).** Software project success is a multi-faceted concept that has been discussed, debated and researched for decades (Kandengwa & Khoza, 2021). The earliest definitions of project success emphasised the importance of project efficiency. Project efficiency refers to whether the project was completed on time, within scope and cost estimations. These criteria, coined the iron triangle trio, have been the de facto standard for decades (Serrador & Turner, 2015). However, this measurement has been criticized as short-sighted as it ends prematurely, meaning the project's success is evaluated when the project management process completes (Thomas & Fernández, 2008). Therefore, the definition of project success needs to expand beyond the project management success criteria because a project can still be successful despite poor project management performance.

Considering project success criteria used across numerous studies as candidates for measuring Agile project success for this research added a layer of rigor. After conducting a review of 28 relevant studies, it became evident what the most cited project success criteria are, namely **S** (Scope), **T** (Time), **C** (Cost), **K** (Customer Satisfaction), **B** (Business Goals), and **Q** (Quality). Due to page count restrictions, these dependent variables are defined in Appendix J. Measuring the 21 SLR-derived factors with each of the six project success criteria seems appropriate, as a previously unmeasured relationship of significance could be unearthed.

**Agile Project Success Factors (Independent Variables).** Agile project success depends on the independent variables included in five themes: Technical Factors, Process Factors, Organisational Factors, People Factors, and Project Factors (Hanslo & Tanner, 2026). These factors, synthesised and described by Hanslo and Tanner (2026), are grounded in CSF theory, which emphasises identifying and prioritising critical factors that are key to a successful outcome.

## RESEARCH METHODOLOGY

**Research Design.** The research design of this study is guided by Saunders et al. (2009) proposed "Research Onion" framework. In this study, a positivist philosophical stance is taken. This decision is consistent with the nature of the research question, the design of the quantitative study, and the methodical process used to create the survey instrument. Ideally suited for identifying significant relationships, this research primarily employs a deductive approach, formulating hypotheses and testing them for statistical significance.

**Research Strategy and Method Choice.** This study employs a quantitative survey design, aligning with the need to collect data from a large sample of practitioners. Through structured questionnaires, surveys systematically collect data on a population's characteristics, practices and perceptions (Kelley et al., 2003). For this study, a mono-method quantitative approach is suitable to systematically gather, analyse, and interpret the data to answer the research question.

**Research Instrument.** The survey instrument was divided into nine sections (Appendix A). Validated measurement items were primarily adapted from previous studies to develop the questionnaire while ensuring they reflected the context of this study's constructs (Appendix B). Participants were asked to reflect on one of their most recently completed Agile projects, rating the corresponding research constructs on a 7-point Likert scale. The validity and reliability of the measurements are what determine the quality of the measuring tool employed (Kimberlin & Winterstein, 2008).

**Sampling Strategy.** The study aimed to identify and explain the CSF contributing to Agile Software Development (ASD) project success as perceived by Agile practitioners. The target population for this study was global Agile practitioners. An Agile practitioner is a member of the Agile team who uses Agile practices and principles to finish a software development project. Due to the geographically dispersed nature of the target population and imprecise knowledge of the population size, the non-probability purposive sampling method was deemed the most appropriate (Etikan & Bala, 2017). Participants were recruited through LinkedIn, a professional networking platform that allows efficient access to a pool of potential respondents. The final sample size for the quantitative survey was 208 Agile practitioners.

**PLS Path Modelling Approach.** PLS-SEM, also called Partial Least Squares Path Modelling (PLS-PM), is used to measure relationships between unobservable Latent Variables (LVs) and measurable Manifest Variables (MVs) (Henseler, 2010). PLS-SEM was selected as the preferred approach for this study. The justification for choosing PLS-SEM over Covariance-based SEM (CB-SEM) is based on the characteristics of the data and the model, such as a focus on prediction, the complexity of the developed model, and its flexibility with non-normal data, whereas CB-SEM is applied in testing or confirming theory and minimises the discrepancy between estimated and sample covariance matrices (Hair & Alamer, 2022). In this study, the quality of the measurement (outer) models' constructs is assessed by analysing the indicator reliability, internal consistency reliability, discriminant validity, and convergent validity. Path coefficients are used to estimate the strength and direction of these relationships. Through analysing the path coefficients and corresponding t-values using bootstrapping, this study tests the hypotheses and the relationships between the exogenous and endogenous constructs of the structural (inner) model.

**Ethical Considerations.** Strict ethical guidelines were followed to protect participants' privacy. Participation in the online survey was voluntary, and anonymity was ensured, as no personally identifiable information was gathered. The research institution's research ethics committee provided ethical approval before data collection.

## RESEARCH RESULTS

The survey respondents (n=208) represented organisations spanning eight global regions. Africa accounted for 55% of the primary locations of the respondents' organisations, while North America, Europe and Asia accounted for 15%, 14% and 8%, respectively. The rest of the responses were spread across Australia and Oceania (3%), Central America (2%), the Middle East (2%), and South America (1%). The most common role among respondents was Software Developer (34%), followed by Scrum Master (20%), Software Architect (8%), Quality Assurance (7%), and Project Manager (6%). The most popular Agile methodology among respondents was Scrum (66%). About 40% of the respondents completed more than 10 ASD projects, and 38% had 6-10 years of Agile software development experience.

### Exploratory Factor Analysis (EFA)

To analyse the survey data, we first conduct an EFA to identify the underlying factors within the dataset. Appendix C provides the scale validity results, with the scree plot and factor loadings. These EFA loadings resulted in 21 constructs (18 factors and three success criteria). Appendix D depicts the scale item loadings after EFA validation. The blue sections highlight the deviations from the TF. After the EFA validation, the factors are grouped into constructs and compared to the hypothesised factors Appendix D.

### Assessing the Measurement (Outer) Model

The outer model is evaluated through Confirmatory Composite Analysis (CCA) using metrics such as outer loadings, indicator multicollinearity, internal consistency reliability, convergent validity, and discriminant validity to ensure the accuracy and validity of the constructs. All outer loadings produced from SmartPLS in this investigation exceed 0.6 and are higher than the minimal value of 0.4 (Hair et al., 2016). Most of the first order constructs demonstrated satisfactory reliability, achieving Cronbach's Alpha and Composite Reliability (CR) scores above 0.70 (Hair et al., 2011). The study's first order constructs all had Average Variance Extracted (AVE) values over 0.50, indicating that the constructs' convergent validity was satisfactory (Fornell & Larcker, 1981). Furthermore, discriminant validity for each of the first order constructs in this investigation was verified using the Fornell-Larcker criterion and Heterotrait-Monotrait Ratio (HTMT). Appendix E details the outer models' statistics.

## Structural (Inner) Model Fit

The inner model is assessed using adjusted R-squared ($R^2$), Standardized Root Mean Squared Residual (SRMR), predictive relevance ($Q^2$), and effect size f-square ($f^2$) to determine the model's overall goodness-of-fit. The Structural (or Inner) Model defines the relationships between unobservable LVs, whereas the outer Model links these LVs to their respective measurable MVs (Hair & Alamer, 2022). Appendix F details the inner models' statistics.

**Coefficient of Determination ($R^2$).** The first order constructs account for 58.2% of the variation in Stakeholder Satisfaction (adjusted $R^2$ of 0.582), 34.2% of the variation in Schedule (adjusted $R^2$ of 0.342), and 22.5% of the variation in Cost (adjusted $R^2$ of 0.225). The conclusion can be drawn that the model fit altogether was good based on the variance accounted for by the endogenous variables.

**Standardised Root Mean Squared Residual (SRMR).** The SRMR in the current model is 0.071. This number indicates a good fit because it is less than the conventional cut-off of 0.08 (Hu & Bentler, 1999).

**Predictive Relevance ($Q^2$).** The $Q^2$ values for the endogenous constructs were, Stakeholder Satisfaction (0.44 - Strong predictive relevance), Schedule (0.17 - Moderate predictive relevance), and Cost (0.05 - Weak predictive relevance) (Hair & Alamer, 2022). A higher positive $Q^2$ value indicates the PLS-SEM model has a smaller prediction error and, therefore, provides higher predictive performance than simply using mean values. For the model to be deemed relevant, $Q^2$, a measure of predictive relevance, should ideally be more than zero (Shmueli et al., 2019).

**Effect Size f-square ($f^2$).** Effect sizes can be classified as big ($f^2 > 0.35$), medium ($f^2$ between 0.15 and 0.34), and small ($f^2$ between 0.02 and 0.14). Notable effect sizes found in this study include:

- **Stakeholder Satisfaction.** Influenced by Team Effectiveness ($f^2$=0.146), Project Management ($f^2$=0.146), and Level of Agile Use ($f^2$=0.023).
- **Schedule.** Impacted by Project Management ($f^2$=0.044) and Project Governance ($f^2$=0.042).
- **Cost.** Influenced by Project Governance ($f^2$=0.040).

## Hypothesis Testing

Following the validation of the measurement model, a rigorous strategy was implemented to examine the theoretical underpinnings via structural model assessment. This involved using 5,000 bootstrap resamples to estimate standard errors and evaluate the significance of direct pathways (Hair et al., 2011). 54 hypothetical assumptions are tested (18 factors for each of the three criteria). The hypotheses testing results identified Project Governance (PG), Project Management (PM), and Team Effectiveness (TE) as having a statistically significant effect on project success in terms of Project Cost, Schedule and Stakeholder Satisfaction. Appendix G details the hypothesis testing statistics.

**Cost.** PG had a significant positive effect ($p < 0.05$) on Cost (coefficient=0.234, t=2.434, p=0.015). Gregory et al. (2016) suggested that further studies be done on this factor, seeing its potential to contribute to project success. Within this study, PG is a significant contributor to ASD project success in terms of project Cost.

**Schedule.** PG (coefficient=0.221, t=2.471, p=0.014) and PM (coefficient=0.285, t=2.141, p=0.032) both have significant ($p < 0.05$) positive effects on Schedule. PG is posited in other studies as an important

Agile project success factor. In this study, PG and PM significantly contribute to Agile project success regarding the project Schedule.

**Stakeholder Satisfaction.** TE (coefficient=0.397, t=2.972, p=0.003) and PM (coefficient=0.413, t=4.626, p=0.000), respectively, exhibit high ($p < 0.01$) and very high ($p < 0.001$) statistically significant positive effects on Stakeholder Satisfaction. Of the 44 synthesised papers (Hanslo & Tanner, 2026), none had TE as a factor. Therefore, this construct is novel when compared to these studies. For this study, TE and PM are significant predictors of Agile project success regarding Stakeholder Satisfaction.

The final model constructs with the latent variable loadings, the path coefficients, and adjusted r-squared (R2) values are shown in Appendix G. The 18 exogenous constructs are color-coded to make it easier to identify their metric values, and the latent variable relationships are displayed with their respective measurement items.

## DISCUSSION OF FINDINGS

The Agile project success TF has gone through three iterations, as depicted in Appendix H. For readability, the final iteration of the TF is depicted in Figure 1. The results suggest that three dependent constructs, Cost (C), Schedule (S), and Stakeholder Satisfaction (SS) showed weak, moderate, and strong predictive relevance. Four independent constructs were identified as enhancing the prediction of these dependent criteria: Level of Agile Use (LU), Project Governance (PG), Project Management (PM), and Team Effectiveness (TE). Three of the 18 validated factors significantly affect ASD project success: PG, PM, and TE. Hypothesis testing further revealed five significant relationships: PG with C and S, PM with S and SS, and TE with SS. Therefore, assessing the measurement model and evaluating the structural model fit ensured that the model is robust and the hypothesised relationships between the constructs could be tested and supported.

### Agile Software Development Project Success Criteria

**Cost (C).** In essence, project **C** as a measure, is correlated with project success. The findings suggest that **C** is still a relevant criterion that needs to be included and evaluated as part of the success criteria for ASD projects.

**Schedule (S).** Schedule as a factor loading effectively integrates a project's time and scope elements. The Project Management Institute (PMI) defines a **S** as "a model for executing the project's activities, including durations, dependencies, and other planning information" (PMI, 2021, p. 58). Furthermore, step one of **S** planning predictive approaches is to "Decompose the project scope into specific activities" translating the project's scope, into a detailed timetable (PMI, 2021, p. 58). This timetable details the planned start and finish times for each task, activity, and milestone (PMI, 2021). The literature supports the importance of **S** as a success criterion and its use as an endogenous construct, in that it integrates and aligns the project's scope and time constraints.

**Stakeholder Satisfaction (SS).** PMI (2021, p. 31) defines stakeholders as "… individuals, groups, or organisations that may affect, be affected by, or perceive themselves to be affected by a decision, activity, or outcome of a portfolio, program, or project." The definition of **SS** in this study is derived from PMI (2021, p. 92) stating that "Stakeholders accept and are satisfied with project deliverables". The construct loadings for **SS** make logical and theoretical sense and are validated by the model and supported by previous literature (Anjani et al., 2021).

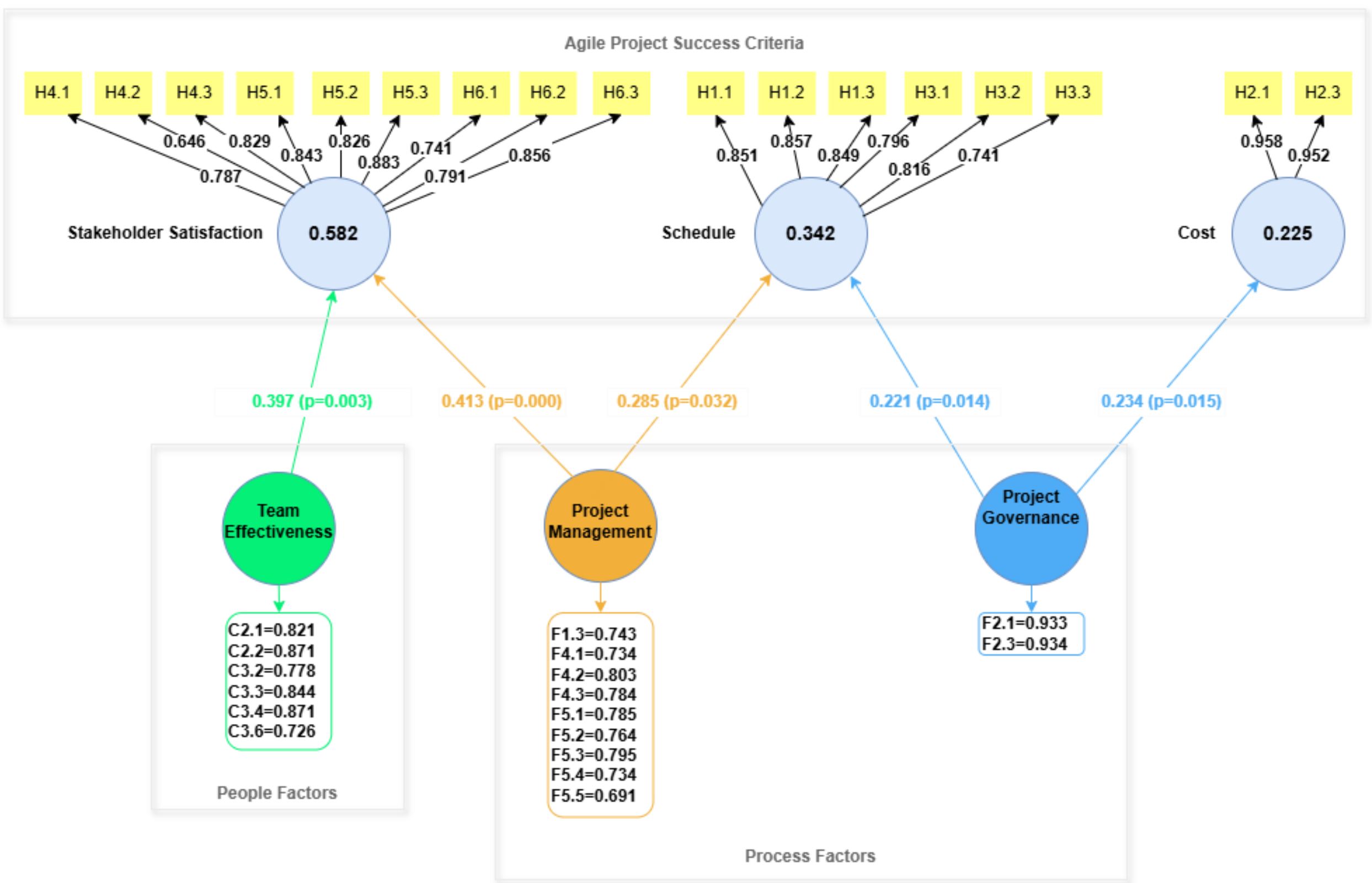

**Figure 1. TF third iteration after hypothesis testing**

## Significant Critical Success Factors

**Hypothesis 3 proposed that a significant positive relationship exists between Level of Agile Use (LU) and Agile Project Success.** The **LU** signifies the degree to which a project embodies the Agile mindset, such as prioritising adaptability, customer collaboration, and functional software delivery over rigid plans and contractual obligations. To clarify, **LU** did not have a statistically significant relationship with Agile project success. Neither did 14 other hypothesised constructs (Appendix I), and therefore they will not be included in the discussion. However, **LU** has a small effect size on SS. Several studies have highlighted the contingent nature of Agile's effectiveness, emphasising the importance of contextual factors and organisational readiness in determining its success (Moloto et al., 2021). In addition, the findings of Ciric Lalic et al. (2022) indicate that projects that are managed in a more Agile way have a positive impact on project success regarding the impact on team satisfaction and preparation for the future, aligning with the PLS-SEM model's effect size of **LU** on SS. The findings suggest that implementing Agile alone is not enough to improve project outcomes.

**Hypothesis 10 proposed that a significant positive relationship exists between Team Effectiveness (TE) and Agile Project Success.** In the framework of this study, **TE** highlights how crucial individual skills and group dynamics are to project success. This finding indicates that a large proportion of variance in SS is explained by **TE**. **TE** exhibits high ($p < 0.01$) significance with SS at a 99% confidence level. Moreover, the PLS-SEM model of this study is good at predicting SS. In essence, **TE**, as a significant exogenous construct, is doing a good job explaining some of the variance in SS and is robust at predicting levels of SS. The theoretical implications of our findings are compelling. **TE** as a CSF and its significant positive relationship with SS suggest from a CSF theory lens that management should pay careful and

consistent attention to this relationship to ensure optimal results. Therefore, organisations need to prioritise and nurture **TE** to improve the chances of delivering quality software products that meet SS.

**Hypothesis 17 proposed that a significant positive relationship exists between Project Governance (PG) and Agile Project Success. PG** is an organised system of oversight and control mechanisms designed to efficiently manage risks and obligations, ensuring that projects adhere to established policies, procedures, and regulations (Sithambaram et al., 2021). **PG** was found to have a statistically significant ($p < 0.05$) positive direct effect on both C and S. **PG** is the only significant factor for C. As a significant construct for S, **PG** explains some of the moderate variance with moderate predictive relevance. **PG** through the CSF theory lens is identified as a significant contributor to Agile project success in terms of C and S, and if done right, can lead to business objectives being met and contribute to organisational success (Rockart, 1979). Therefore, this study provides strong evidence for **PG**'s role in improving efficiency, predictability, and innovation by creating the ideal conditions for Agile projects to succeed (Onik et al., 2017).

**Hypothesis 18 proposed that a significant positive relationship exists between Project Management (PM) and Agile Project Success.** The project's definition and planning are part of the **PM** process (PMI, 2021). In essence, **PM** is about doing the work (focusing on the execution), while PG is about overseeing the work (focusing on the oversight). As such, the items that have merged for this construct are about doing the work. **PM** has a significant ($p < 0.05$) positive direct effect on S and a very high significant ($p < 0.001$) positive direct effect on SS at a 99.9% confidence level. Where this finding adds to the body of knowledge is advancing our understanding of the broader definition of **PM** with its expanded aspects such as the project definition, prioritising its development requirements, planning, and incorporating quality assurance techniques such as early testing and change control management. This finding proposes that executing the project through the suggested **PM** process will improve your chances of satisfying your stakeholders and adhering to the schedule. From a theoretical perspective **PM** serves as a CSF of Agile software development projects. **PM**, therefore, is posited to lead to improved SS and S adherence by ensuring that projects are well-defined, planned for, and executed correctly.

## Comparison with Previous Research

It is crucial to highlight that, to the best of our knowledge, previous academic studies could not provide as few factors and relationships as this study. Specifically, these prior studies examined a wide array of variables contributing to ASD success, encompassing people, process, technology, project, and organisational factors. It is posited that this study's findings help to simplify the requirements for organisations to execute a successful Agile project. It is worth mentioning that the findings of the private research and advisory firm specialising in IT project performance (the Standish Group) in their *CHAOS 2020: Beyond Infinity* report (StandishGroup, 2020) had some resemblance to this study's findings. For instance, they identified three key practices (critical factors) for project success: 1) a good sponsor, 2) a good team, and a 3) good place. The report further articulates how project success is measured in six attributes: 1) on-time, 2) on-budget, 3) on-target, 4) on-goal, 5) value, and 6) customer satisfaction. What we can take away from the Standish Group 2020 report is that there are a few factors that organisations should get right to ensure organisational performance, aligning with the CSF theory. However, unlike this study, the Standish Group's full investigation is not made available for researchers to review, leaving knowledge gaps about the methodology and the instrument used for collecting data. In addition, there is no explanation of how the sample of participants was selected (Varajão & Trigo, 2024). Therefore, their scientific reporting is lacking.

## CONCLUSION

**Summary of the Findings.** The findings suggest that as few as three constructs, namely Team Effectiveness (People factor), Project Management (Process factor), and Project Governance (Process factor), through five relationships, contributes to predicting ASD project success as measured in terms of Stakeholder Satisfaction, Cost, and Schedule adherence. From the CSF theory perspective, these three CSF and five unique relationships are the key areas organisations and Agile teams must prioritise for positive outcomes (Rockart, 1979).

**Limitations and Future Research.** The sampling strategy and the sample size of 208 present potential limitations. Further studies could replicate the positivist survey and PLS-SEM modelling used in this study. Ideally, the sample size should be increased and distributed more evenly across the under-represented regions. Additional research could implement inductive approaches to investigate how these critical factors contribute to Agile software development project success. Finally, a future study could develop new scale items for emerging constructs, such as AI augmentation, virtual team competence, and AI governance.

**Implications and Conclusion.** Based on the findings, organisations should prioritise these three factors and five relationships to ensure successful project outcomes. The PLS-SEM analysis validated a model aligned with CSF theory, isolating the critical few factors that drive Agile software development project success. Compared to previous research, this offers an elegant level of modelling simplicity, highlighting that organisations should purposefully direct their attention toward these specific areas.

## SUPPLEMENTARY MATERIAL

The survey instrument, detailed statistical outputs, and additional framework diagrams are available in the online appendices at: https://osf.io/b932g/overview?view_only=bc18100a3b9048db9cac2d9f8d55ce8b